\documentclass{aa}
\usepackage{graphicx,epsfig,fancyhdr,rotating,amsmath,natbib}
\usepackage[varg]{txfonts}
\usepackage{epsfig,rotate}

\newcommand{\degree}{\ensuremath{^\circ}}

\begin{document}

\title{Exploring the damping of Alfv\'en waves along a long off-limb coronal loop, up to 1.4 R$_\odot$}
\subtitle{}

\author{Girjesh~R. Gupta\inst{\ref{inst1},\ref{inst2}} \and G. Del Zanna\inst{\ref{inst1}} \and H.~E. Mason\inst{\ref{inst1}}}

\institute{DAMTP, Centre for Mathematical Sciences, University of Cambridge, Wilberforce Road, Cambridge CB3 0WA UK \label{inst1}
\and
Udaipur Solar Observatory, Physical Research Laboratory, Badi Road, Udaipur 313 001, India \email{girjesh@prl.res.in} \label{inst2} }

\date{}

\abstract{The Alfv\'en wave energy flux in the corona can be explored using the electron density and velocity amplitude of the waves. 
The velocity amplitude of Alfv\'en waves can be obtained from the non-thermal velocity of spectral line profiles. 
Previous calculations of the Alfv\'en wave energy flux with height in
active regions and  polar coronal holes have provided
evidence for the damping of Alfv\'en waves with height.
We present off-limb Hinode Extreme-ultraviolet Imaging Spectrometer (EIS) observations of a long coronal loop up to 1.4~R$_\odot$.
We obtained the electron density along the loop and found 
the loop to be almost in hydrostatic equilibrium. We obtained
the temperature using the emission measure-loci (EM-loci) method and found the loop to be
isothermal across, as well as along, the loop with a temperature of 
about 1.37 MK. We significantly improve the estimate of non-thermal velocities over previous studies by using the estimated ion
(equal to electron) temperature. Estimates of electron densities are improved using the significant updates of the CHIANTI v.8 atomic
data. More accurate measurements of propagating Alfv\'en wave energy along the coronal loop and its damping are presented up to
distances of 1.4 R$_\odot$, further than have been previously explored. The 
Alfv\'en wave energy flux obtained could contribute to a significant part of the coronal losses due to radiation along the loop.}


\titlerunning{Damping of Alfv\'en waves along a long off-limb coronal loop}
\authorrunning{Gupta et al.}

\maketitle

\section{Introduction}
\label{sec:intro}

The definitive physical processes responsible for the heating of solar corona and acceleration of solar wind have not yet been
identified. There are several theories which exist to explain phenomena such as wave dissipation (AC) models, foot-point stressing
(DC) models, turbulence models, and Taylor relaxation models \citep[see recent review by][]{2018arXiv181100461C}. 
Coronal loops are the basic building blocks of the solar corona. \citet{1998Natur.393..545P} found that physical parameters such as
the temperature profile along the coronal loop are highly sensitive to heating mechanisms. Thus, to distinguish among the
different heating models, accurate measurements of basic plasma parameters along the coronal loops, such as temperature,
density, filling factor, velocity, non-thermal velocity, and magnetic field are essential \citep[e.g.][]{2015ApJ...800..140G,2017ApJ...842...38X}.
Several observational studies of loop characteristics, such as the derivation of temperatures and densities, have been  published
\citep[see e.g.][]{2003A&A...406.1089D,2004ApJ...608.1133L,2008ApJ...680.1477A,2009ApJ...694.1256T,2014ApJ...780..177L,2017ApJ...835..244G}.
More details can be found in recent review by \citet{2018LRSP...15....5D}.

In the wave dissipation model of the solar atmosphere convective motions at the magnetic foot-points of the loop 
are assumed to generate wave-like fluctuations. This wave energy is then transmitted up into the corona, where
the conversion to heat can occur \citep[e.g.][]{2007ApJS..171..520C,2011ApJ...736....3V}. Among the different types of waves, Alfv\'en waves
are the least damped during propagation to the chromosphere and carry sufficient energy to heat the solar corona
\citep[see reviews by][]{2012RSPTA.370.3193D,2013SSRv..175....1M}.

However, in the corona, several spatially unresolved structures may be present along the line of sight. 
These unresolved structures may be oscillating with different phases and could lead to excess broadening of spectral line profiles.
Spectroscopic observations of coronal loops have clearly shown that the line widths of observed emission lines are significantly
broadened in excess of their corresponding thermal widths 
\citep[e.g.][]{2007ApJ...667L.109D,2012ApJ...754..153D,2014ApJ...780..177L,2014ApJ...786...28A,2016ApJ...820...63B,2016ApJ...827...99T,2017ApJ...836....4G,2017ApJ...842...38X}.
These observed non-thermal broadenings of spectral line profiles in the corona are expected to be proportional
to Alfv\'en wave amplitudes \citep[e.g.][]{1990ApJ...348L..77H,1998SoPh..181...91D,1998A&A...339..208B,2001A&A...374L...9M}. 
Thus, upon comparison of non-thermal velocity estimates with height,
we can infer the variation of Alfv\'en wave amplitude with height. Using such measurements, evidence of the damping of Alfv\'en waves
has been reported along the coronal loops \citep{2014ApJ...780..177L,2017ApJ...836....4G}.  

In this paper, our main focus is on the estimation of electron density, temperature, plasma filling factor, and non-thermal velocity along
the coronal loop, up to a very large distance. Further, we aim to obtain the total Alfv\'en wave energy flux along the loop so as to find any
signatures of Alfv\'en wave damping. For this purpose, we identified a unique set of spectroscopic data covering a 
very large distance in the off-limb active and quiet Sun regions observed by Extreme-ultraviolet Imaging Spectrometer \citep[EIS;][]{2007SoPh..243...19C} 
on board Hinode \citep{2007SoPh..243....3K}. Details of the observations are described in Section~\ref{sec:data}.
We employ several spectroscopic methods, which are described in Section~\ref{sec:result}  to obtain the electron number density,
temperature, plasma filling factor, effective and non-thermal
velocity, and to derive the Alfv\'en wave energy flux. 
The results obtained are discussed and summarised in Section~\ref{sec:discuss}.

\section{Observations and data analysis}
\label{sec:data}

\begin{figure*}[htbp]
\centering
\includegraphics[width=0.7\textwidth]{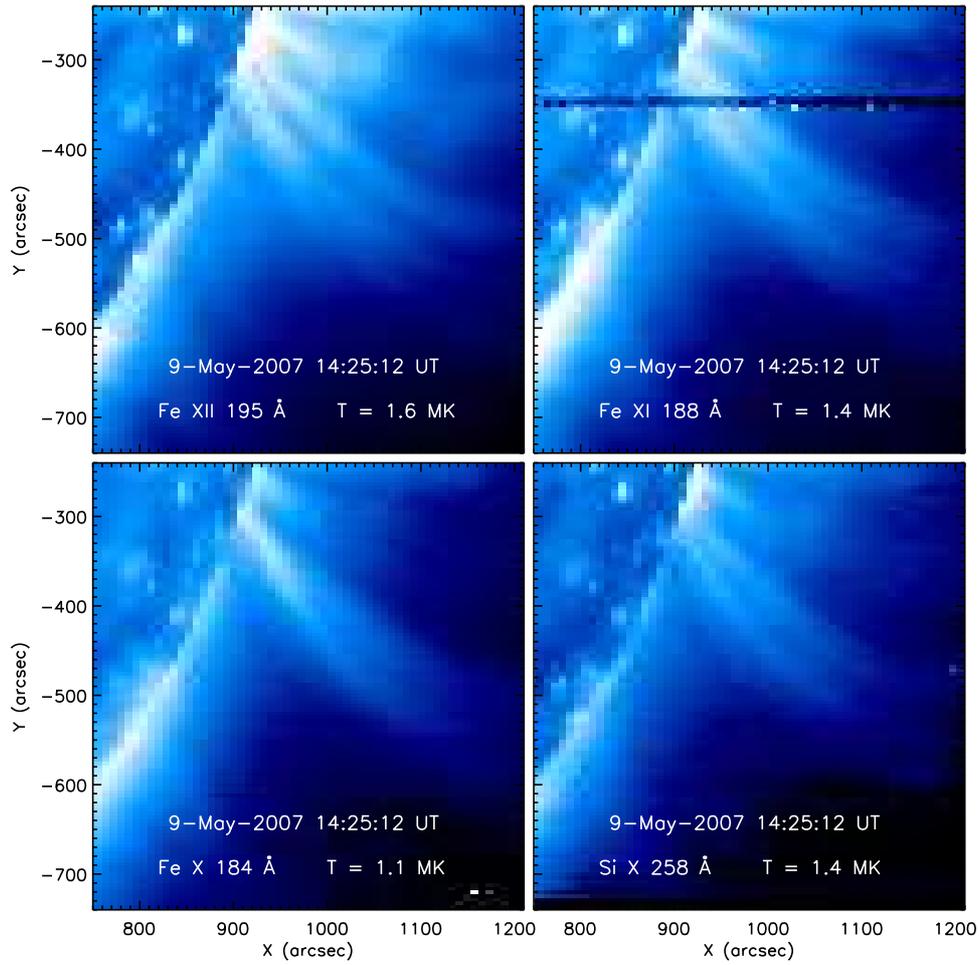}
\caption{Intensity maps obtained from Fe~{\sc x} 184.54, Fe~{\sc xi} 188.23, Fe~{\sc xii} 195, and Si~{\sc x} 258 \AA\  
showing clear long loops in all the panels.}
\label{fig:context}
\end{figure*}

Because of the constraints on Hinode pointing outside the solar limb, off-limb observations of coronal loops reaching far out in the corona
using EIS are very rare with the exception of several observations
reaching far out in the polar regions. In order to overcome this issue and
to reach greater distances, one of us (GDZ) designed an EIS engineering study (study ID 141) to extract spectra from the
bottom half of the long EIS slit. The Joint Hinode Observational Program (Hinode HOP 7) was coordinated to obtain simultaneous 
SOHO/Hinode/TRACE/STEREO observations during May 7--10 2007. The HOP was summarised in \citet{2009ASPC..415..315D} and some results
from other days of the campaign are presented in \citet{2018ApJ...865..132D,Delzanna2019}.
On May 9, a long coronal loop appeared on the west limb. The region
was rastered by the EIS 2'' slit with an exposure time of 60 s. 
The EIS slit moved with an 8'' step size and a total of 60 exposures were taken, which covered the total field of view of $480''\times 512''$.
We selected the raster which began on that day at 14.25.12 UT for a detailed analysis.
We binned the data over four pixels along the slit to improve the signal-to-noise ratio.
We followed the standard solar software (SSW) procedure EIS\_PREP for preparing the EIS data.
All the EIS spectral line profiles were fitted with a Gaussian function using EIS\_AUTO\_FIT \citep{Young2015}.
The routine also provides $1\sigma$ error bars on the fitted parameters. The EIS sensitivity is degrading with time,
therefore we further recalibrated the intensities and related errors using the method of \citet{2013A&A...555A..47D}. 
We assumed the instrumental profile to be Gaussian and subtracted the instrumental width following the
method described in \citet{Young2011}. Moreover, our analyses \citep{Delzanna2019} 
of variation of instrumental width along the slit also show similar results of 65--72 m\AA,\ which is within a
few m\AA\ of \citet{Young2011}.
Spatial offsets in the solar-X and solar-Y directions between images obtained from different
wavelengths were corrected with respect to the image obtained from the Fe XII 195.12 \AA\ spectral line.

\begin{figure*}[htbp]
\centering
\includegraphics[width=0.9\textwidth]{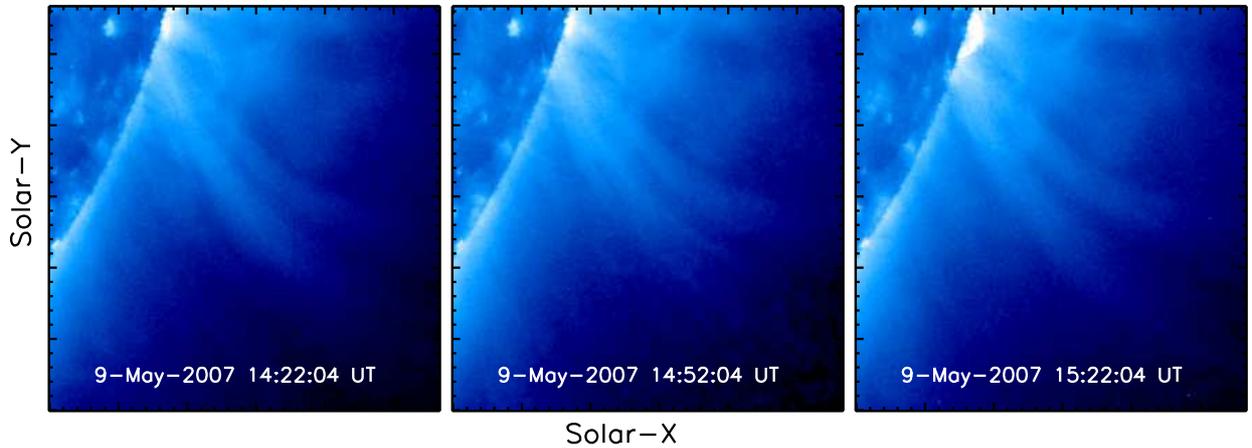}
\caption{Intensity maps obtained from the STEREO B EUVI 195 \AA\ passband showing clear long loops at different times as labelled.}
\label{fig:euvi}
\end{figure*}

In Figure~\ref{fig:context}, we plot monochromatic intensity maps of the off-limb region obtained from Fe~{\sc x} 184.54,
Fe~{\sc xi} 188.23, Fe~{\sc xii} 195, and Si~{\sc x} 258 \AA\ spectral lines. Images obtained from these lines 
show the presence of a clear coronal loop extending very far
off-limb. In the images from other spectral lines, the loop is more difficult
to discern. In Figure~\ref{fig:euvi}, we also show the STEREO-B EUVI images taken in 195 \AA\ passband at different times overlapping
the time interval of EIS raster scan. During this interval, STEREO-B was just 2.3\degree\ ahead of the Sun-Earth line in the
ecliptic plane, thus viewing almost same part of the Sun. Images obtained at different times show the clear presence of two long
loops reaching far out. The loop at the bottom shows dynamic motion, whereas the upper loop is relatively stable. 
Thus, we focus our attention on studying the upper loop.

We calculated the contribution function of the EIS spectral lines chosen for this study using CHIANTI v.8 
\citep{1997A&AS..125..149D,2015A&A...582A..56D} at a constant electron number density $N_e=10^8$ cm$^{-3}$. 
The peak formation temperatures of all the selected lines are labelled in Figure~\ref{fig:context}.
We also identified density-sensitive lines Fe~{\sc xii} 186.88 and Si~{\sc x} 258.37 \AA\ and used these  for the purpose of 
deriving electron number densities along the loop length.

\begin{figure}[htbp]
\centering
\includegraphics[width=0.45\textwidth]{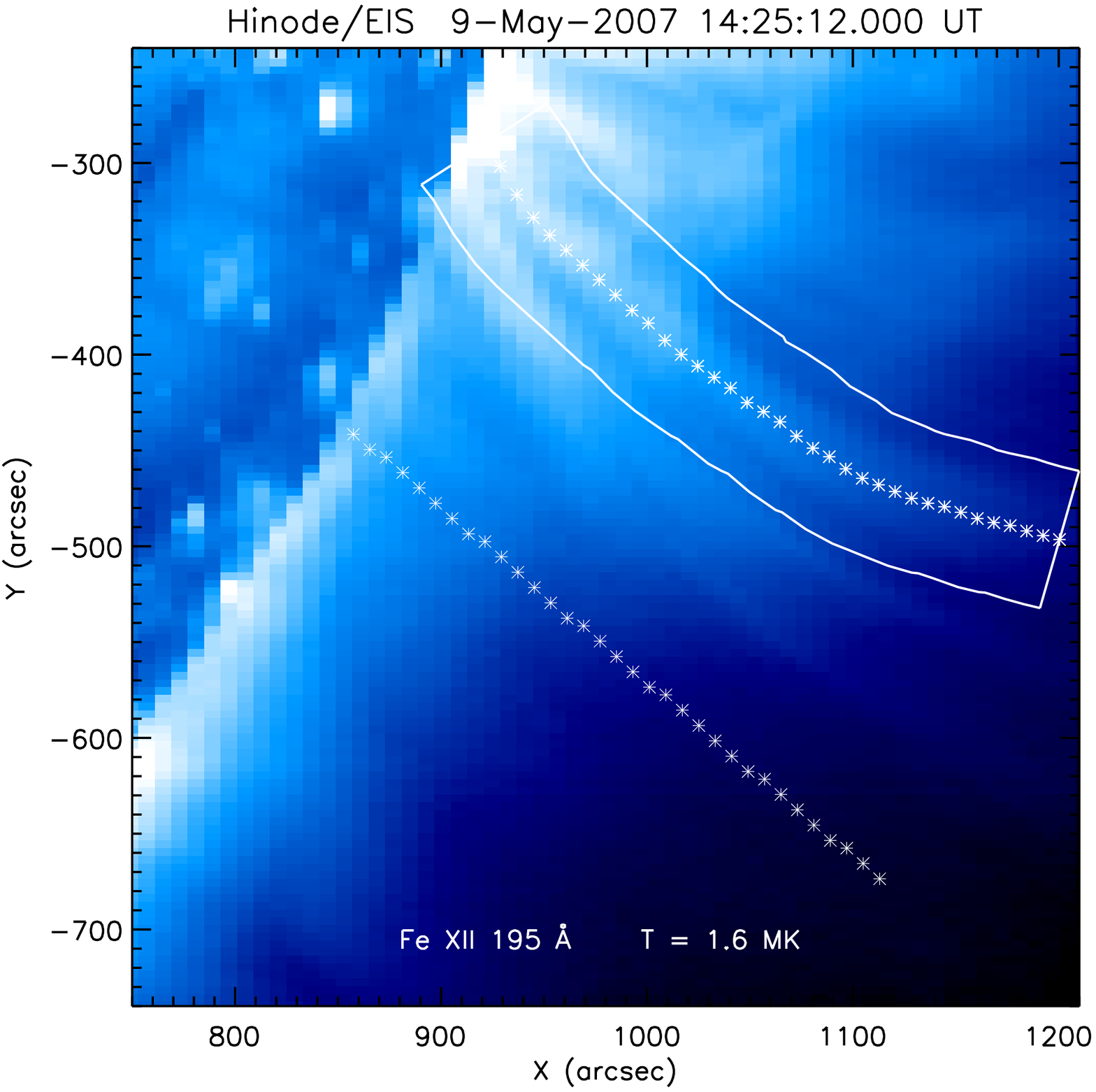}
\caption{Intensity maps obtained from  Fe~{\sc xii} 195 \AA\  showing
  the loop traced out to study in detail.
The quiet Sun region is also traced out for the purpose of determining
the background. The white curved box indicates region chosen along the curved loop.}
\label{fig:context195}
\end{figure}

In Figure~\ref{fig:context195}, we plot monochromatic intensity maps of the off-limb active region obtained from Fe~{\sc xii} 195.12 \AA .
We trace the loop using this image. The traced points are indicated with asterisks (*). We also identified a quiet region near the loop for the
background/foreground subtraction purposes (also marked with asterisks). We created a curved box along the loop to study the loop parameters.
The curved box is shown with a white continuous line.

\begin{figure}[htbp]
\centering
\includegraphics[width=0.45\textwidth]{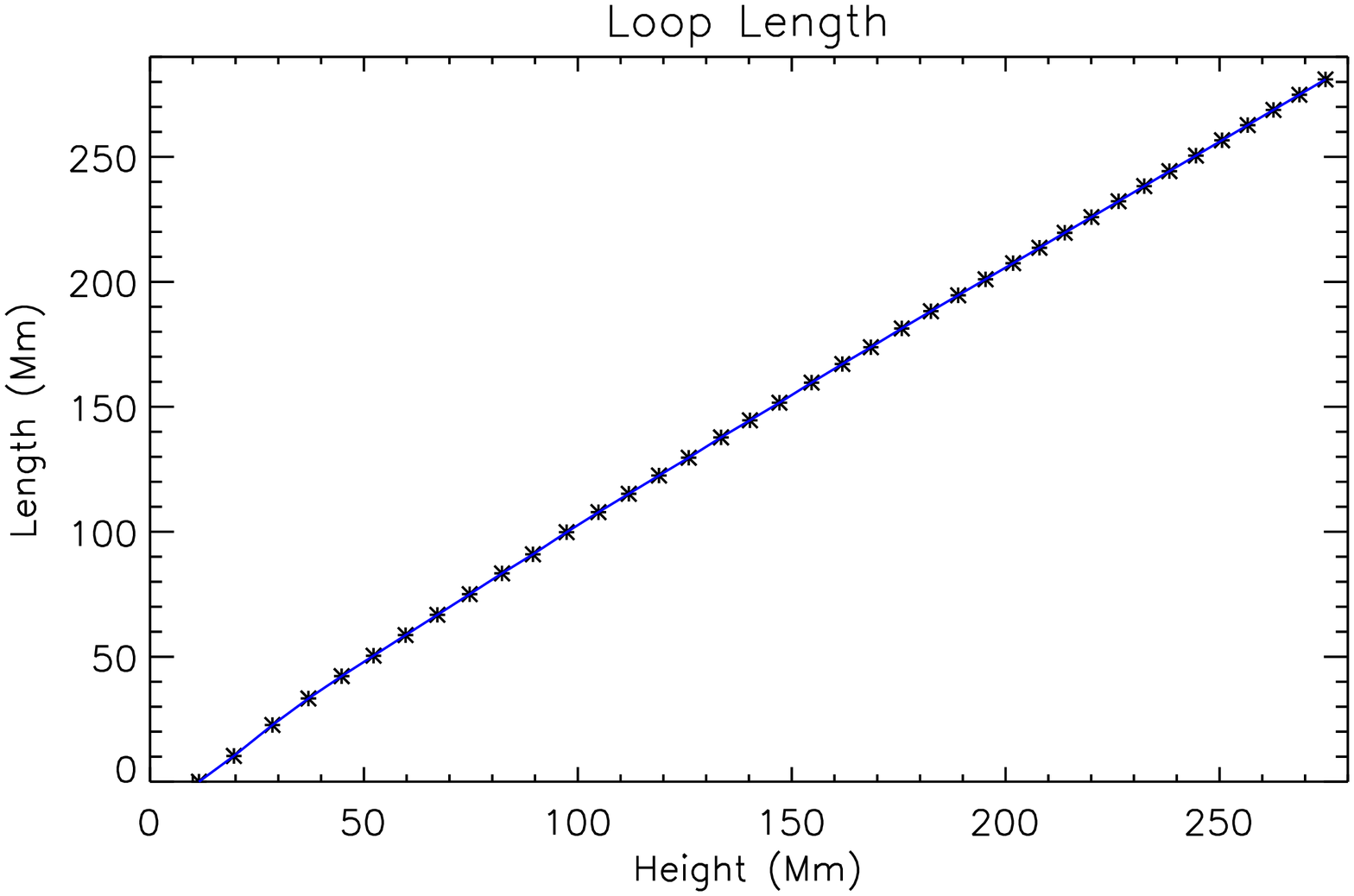}
\includegraphics[width=0.45\textwidth]{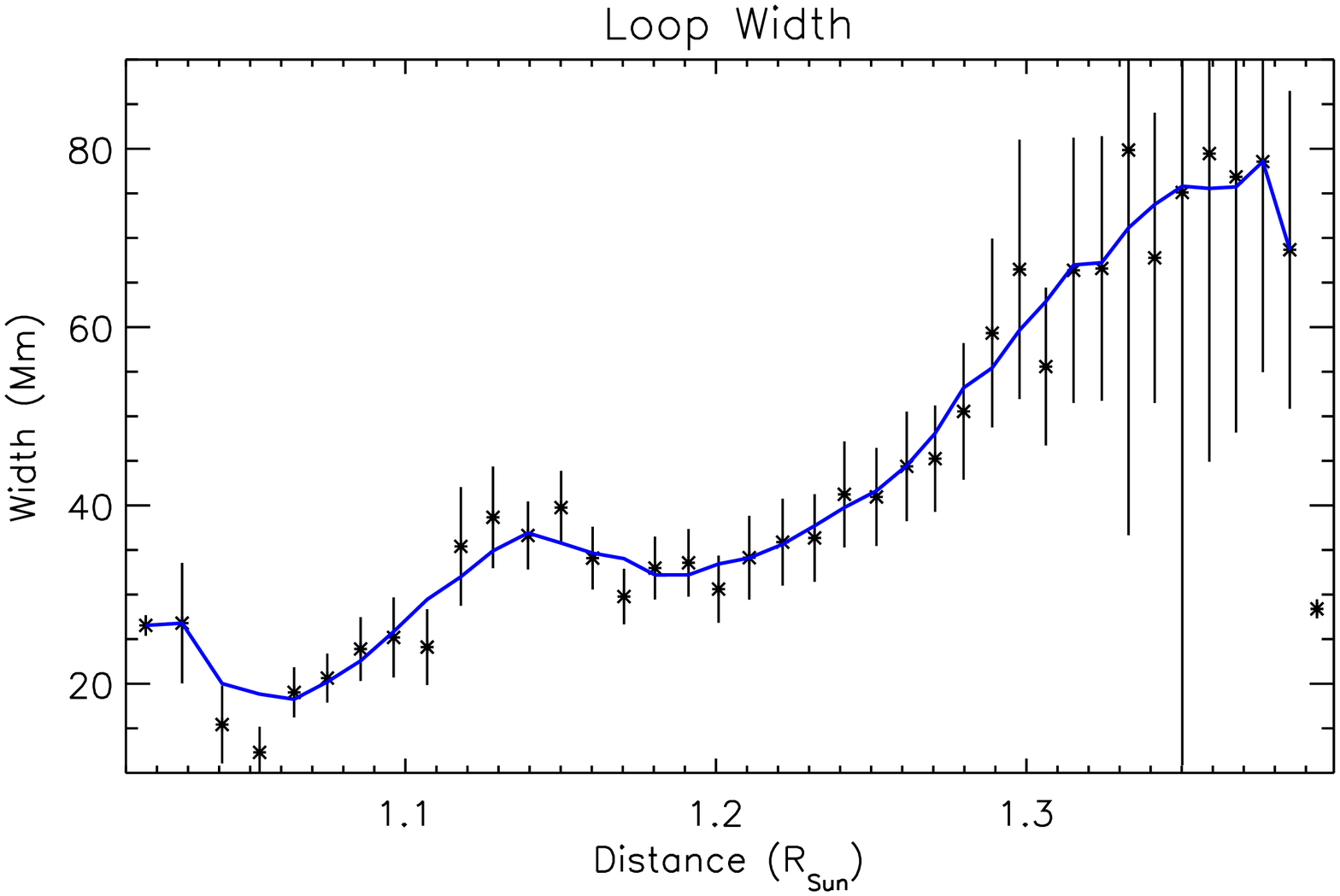}
\caption{Length (top panel) and width (bottom panel) of the long
  coronal loop with distance as obtained from the Fe~{\sc xii} 193 \AA\ intensity map.}
\label{fig:geom193}
\end{figure}

From the curved box, we obtained an intensity across the loop at each
of the traced points using the image from the Fe~{\sc xii} 193 \AA\ line.
 We fitted these intensity profiles across the loop positions with a
 Gaussian function. The full width half maximum (FWHM) of the Gaussian profile provides a geometric
  width of the loop at that position. We obtained the height of the loop with respect to the solar limb. We also summed individual segments of 
  the loop to obtain the length of the loop. The loop length and width obtained are plotted in the top and bottom panels of Figure~\ref{fig:geom193}, respectively. 
The figure clearly shows that the loop is traced up to the height of 270 Mm and has a total length more than 280 Mm. This also indicates that actual length and height
 of the loop would have been even longer and likely went beyond the EIS field of view (FOV). The geometric width of the loop increases with height from 20 Mm near limb to nearly 80 Mm 
 at the distance of 1.37 R$_{\odot}$. 

\begin{figure*}[htbp]
\centering
\includegraphics[width=0.7\textwidth]{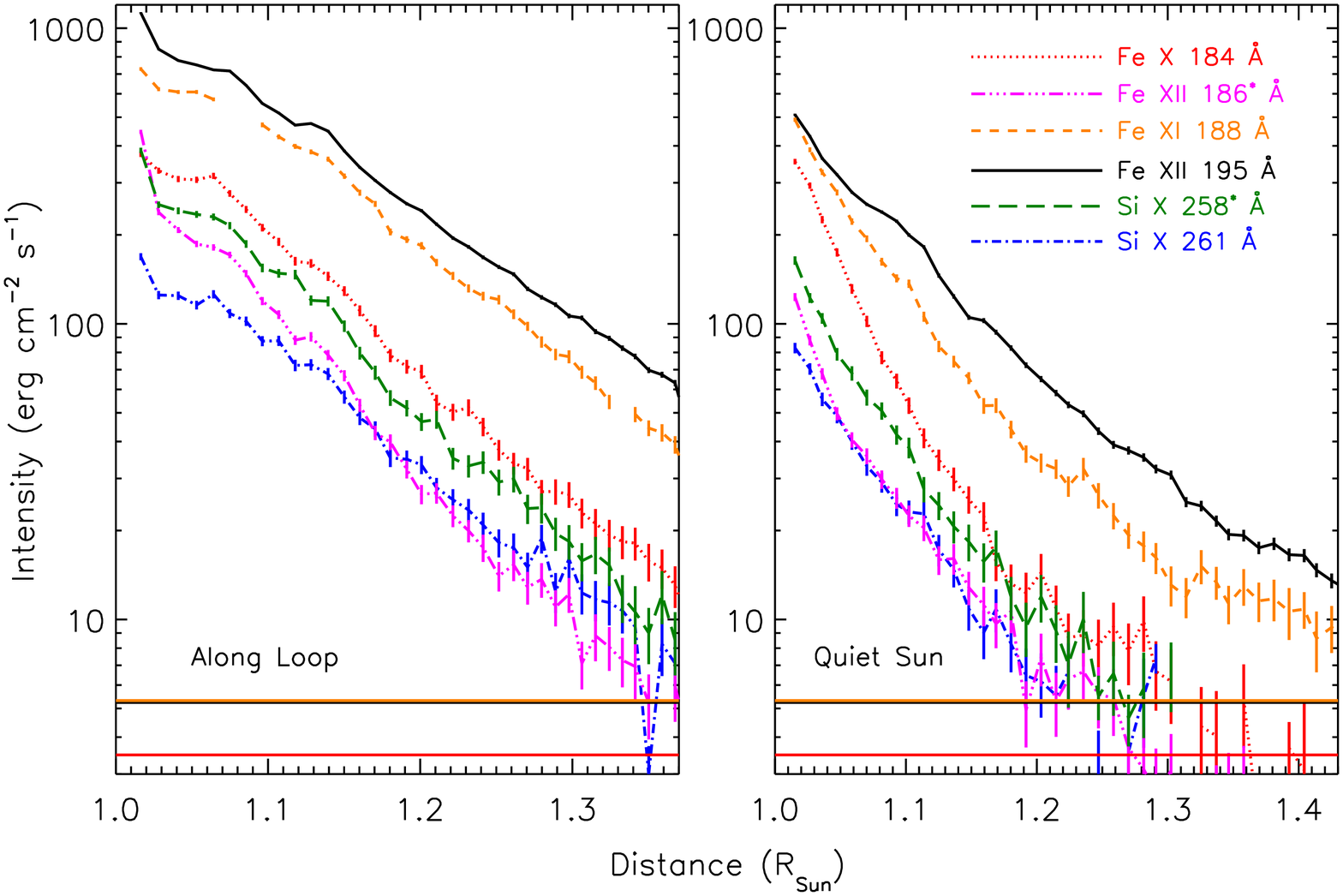}
\caption{Intensity obtained from  Fe~{\sc x} 184.54, Fe~{\sc xii} 186.88$^*$, Fe~{\sc xi} 188.23, Fe~{\sc xii} 195.12, 
and Si~{\sc x} 258.37$^*$, 261.04 \AA\ spectral lines along the coronal loop (left panel) and quiet Sun region (right panel) 
with height. The intensities are recalibrated using the method of \citet{2013A&A...555A..47D}. The asterisks denote respective spectral
lines are density sensitive. Horizontal lines denote expected level of scattered light as per \citet{Ugarte2010}.}
\label{fig:intensity}
\end{figure*}

\section{Results}
\label{sec:result}

We obtained the various parameters along the loop such as intensity, density, temperature, plasma filling factor, 
and Alfv\'en wave energy flux. We describe the estimation of all these parameters in the following subsections.

\subsection{Intensity}
\label{sec:int}

In Figure~\ref{fig:intensity}, we plot the intensities obtained along the loop and quiet Sun region
with height. The 1$\sigma$\ error bars associated with the data points are overplotted.
It is clear from the plots that the spectral lines have a good enough signal to perform a further detailed analysis.

We also estimated the level of scattered light in the intensities using available field of view of
on-disc region as suggested by \citet{Ugarte2010}, which is expected to be about 2\% of the disc intensity.
This level is also overplotted in the Figure~\ref{fig:intensity} for all respective spectral lines.
However, scattered light estimate of \citet{2012ApJ...753...36H}  indicates that a 2\% 
level of scattered light contribution is an overestimation in the polar coronal hole region.
\citet{2018ApJ...865..132D} analysed the same type of observation as performed in this work on the May 10 (one day after).
Their analysis indicates that scattered light usually assumed at 2\% level is an overestimate even up to 1.3 R$_{\odot}$.
Therefore, it is safe to assume that scattered light contribution in the coronal loop is negligible as structure is visible 
until the end of EIS  FOV, thus can conveniently be ignored in the further analysis.

\begin{figure}[htbp]
\centering
\includegraphics[width=0.42\textwidth]{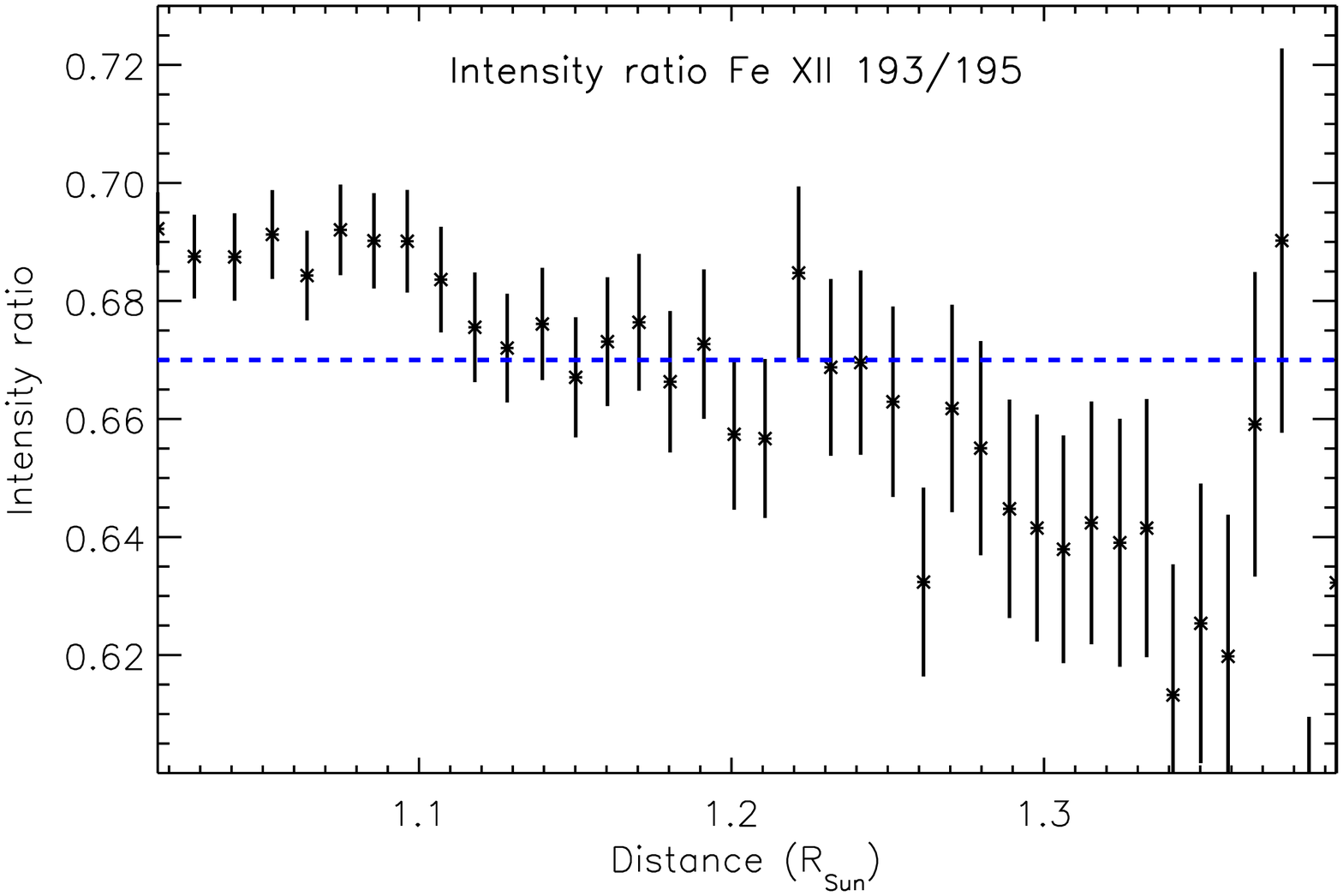}
\caption{Measured intensity ratios of Fe~{\sc xii} 193.51 and 195.12 \AA\  lines along the coronal loop. 
This ratio is expected to be constant at 0.67 as per CHIANTI v.8 database (dashed line).}
\label{fig:int-rat}
\end{figure}

We also obtained the variation of intensity ratios of Fe~{\sc xii} 193.51 and 195.12 \AA\ lines along the loop and 
plot these in Figure~\ref{fig:int-rat}. From the CHIANTI database, this ratio is expected to be constant at 0.67 without
any variation with height. However, recently \citet{Delzanna2019} reported the Fe~{\sc xii} 193.51 and 195.12 \AA\ 
intensity ratios to be different from the expected theoretical value
where the deviation is about 30-40\% in the quiet and active regions.
These authors speculated the cause for these anomalies to be either an opacity effect or some non-linear behaviour in the CCD counts.
However, in our observation, the variation in the intensity ratio is relatively small ($<10$\%).

\subsection{Electron density along the coronal loop}
\label{sec:dens}

\begin{figure*}[htbp]
\centering
\includegraphics[width=0.7\textwidth]{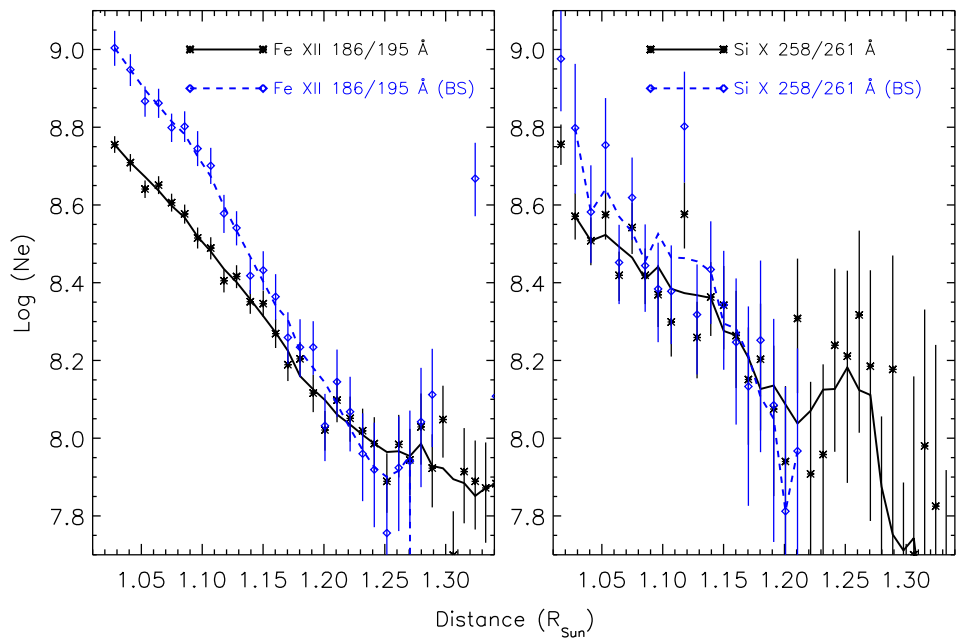}
\caption{Electron density obtained from the Fe~{\sc xii} line ratio method with (dashed lines) and without (solid lines)  BS, as labelled.}
\label{fig:density}
\end{figure*}

\begin{figure*}[htbp]
\centering
\includegraphics[width=0.7\textwidth]{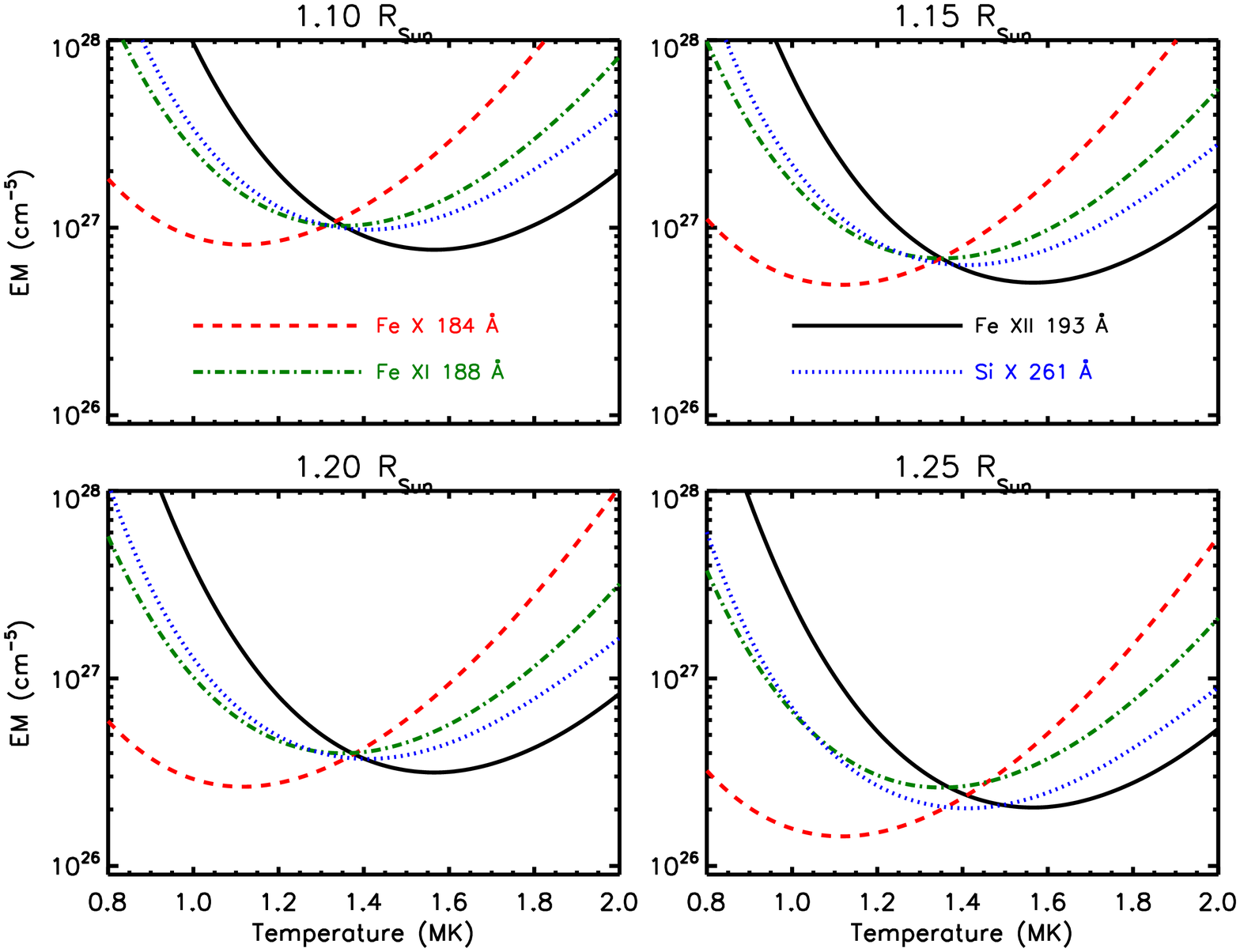}
\caption{Emission measure (EM) loci curves obtained from Fe~{\sc x} 184.54, Fe~{\sc xi} 188.23, Fe~{\sc xii} 193.51, and 
Si~{\sc x} 261.04 \AA\ spectral lines at different heights as labelled.}
\label{fig:em}
\end{figure*}

We estimated the electron number density along the loop using the line ratio method \citep[see recent review by][]{2018LRSP...15....5D}.
For the purpose, we identified the density
sensitive line pairs of Fe~{\sc xii} $\lambda 186.88/ \lambda 195.12$ and Si~{\sc x} $\lambda 258.37/ \lambda 261.04$. 
In Figure~\ref{fig:density}, we plot densities obtained from both the line pairs. We also obtained these densities after subtracting
the background as identified from the quiet region. The densities obtained from the background subtracted intensities are relatively higher
than those obtained without background subtraction (BS). A closer inspection indicates that densities obtained with BS
have higher values compared to densities obtained without BS only near the limb, whereas in the far off-limb region both
the densities have almost the same values. This is because the background signal is present only near the limb region and in 
far off-limb region, there is hardly any background signal. Also, there are huge error bars on the densities obtained from 
the Si~{\sc x} $\lambda 258.37/ \lambda 261.04$ line pair. Electron number densities along the loop as obtained from
Fe~{\sc xii} $\lambda 186.88/ \lambda 195.12$ show the number density
to be around $10^9$ cm$^{-3}$ near the limb, which falls to $\approx 10^{7.9}$ cm$^{-3}$ at a far off-limb distance.

\subsection{Electron temperature along coronal loop}
\label{sec:temp}

\begin{figure}[htbp]
\centering
\includegraphics[width=0.45\textwidth]{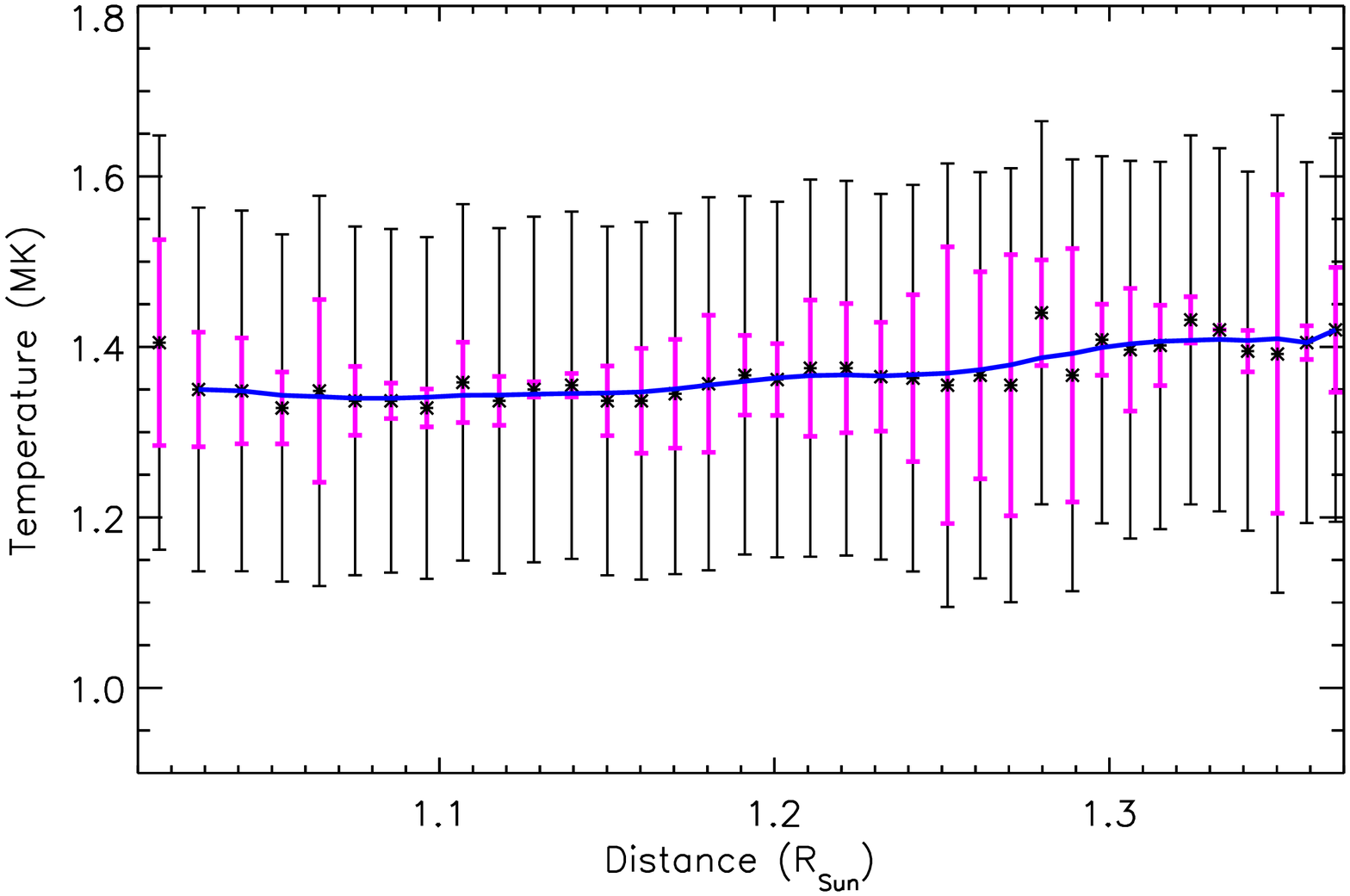}
\caption{Temperature obtained along the coronal loop as obtained from the EM-loci curves.}
\label{fig:temp}
\end{figure}

The electron temperature of the plasma can be estimated using emission lines from ions of different ionisation stages.
Since contribution functions of spectral lines are highly temperature dependent, 
the observed intensity can be converted in to temperature by analysing different ions.
We used the technique of the emission measure-loci (EM-loci) method to estimate the temperature along the coronal loop
\citep[e.g.][]{2002A&A...385..968D}. In this method, EM curves obtained from different ions are plotted as a function of temperature. 
If the emitting region is isothermal then all of the curves would cross at a single location, thus, indicate a single temperature.
We used density insensitive emission lines Fe~{\sc x} 184.54, Fe~{\sc xi} 188.23, Fe~{\sc xii} 193.51, and Si~{\sc x} 261.04 \AA\
to obtain EM-loci curves. Thus, we obtained EM-loci curves at all locations along the loop to obtain electron temperatures along the loop
\citep[see e.g.][]{2015ApJ...800..140G}. In Figure~\ref{fig:em}, we show EM curves obtained only at distance of 1.10, 1.15, 1.20, and
1.25 R$_{\odot}$. All the EM curves cross at almost the same temperature at all locations, suggesting that the loop is nearly
isothermal along and across its length. However, there is a slight difference in temperature crossing points among different lines.
Thus, we choose an average temperature at crossing points as the
electron temperature and standard deviation of temperatures at crossing points 
as standard error bars on the electron temperatures. However, there also exist uncertainties in the atomic data which are
utilised in the CHIANTI database. Although, it is difficult to quantify uncertainties present in various atomic data, in this work
we assume a total of about 15\% uncertainties present in the contribution function of spectral lines. Therefore for error estimation,
we assumed 15\% errors on EMs and associated temperature.  Using this we again calculated larger error bars on the 
temperature along the loop. In Figure~\ref{fig:temp}, we plot electron temperatures obtained along the loop.
Error bars obtained from standard deviation of temperatures at crossing points are plotted in magenta (small error bars),
whereas cummulative error bars obtained after taking in to account all the errors are plotted in black (large error bars).
The figure indicates that the temperature obtained from the EM-loci method is almost constant along the loop within the 
smaller error bars. However, a slight increase in the temperature is also observed at the far distances.

\subsection{Comparison with hydrostatic equilibrium}
\label{sec:hydro}

\begin{figure}[htbp]
\centering
\includegraphics[width=0.48\textwidth]{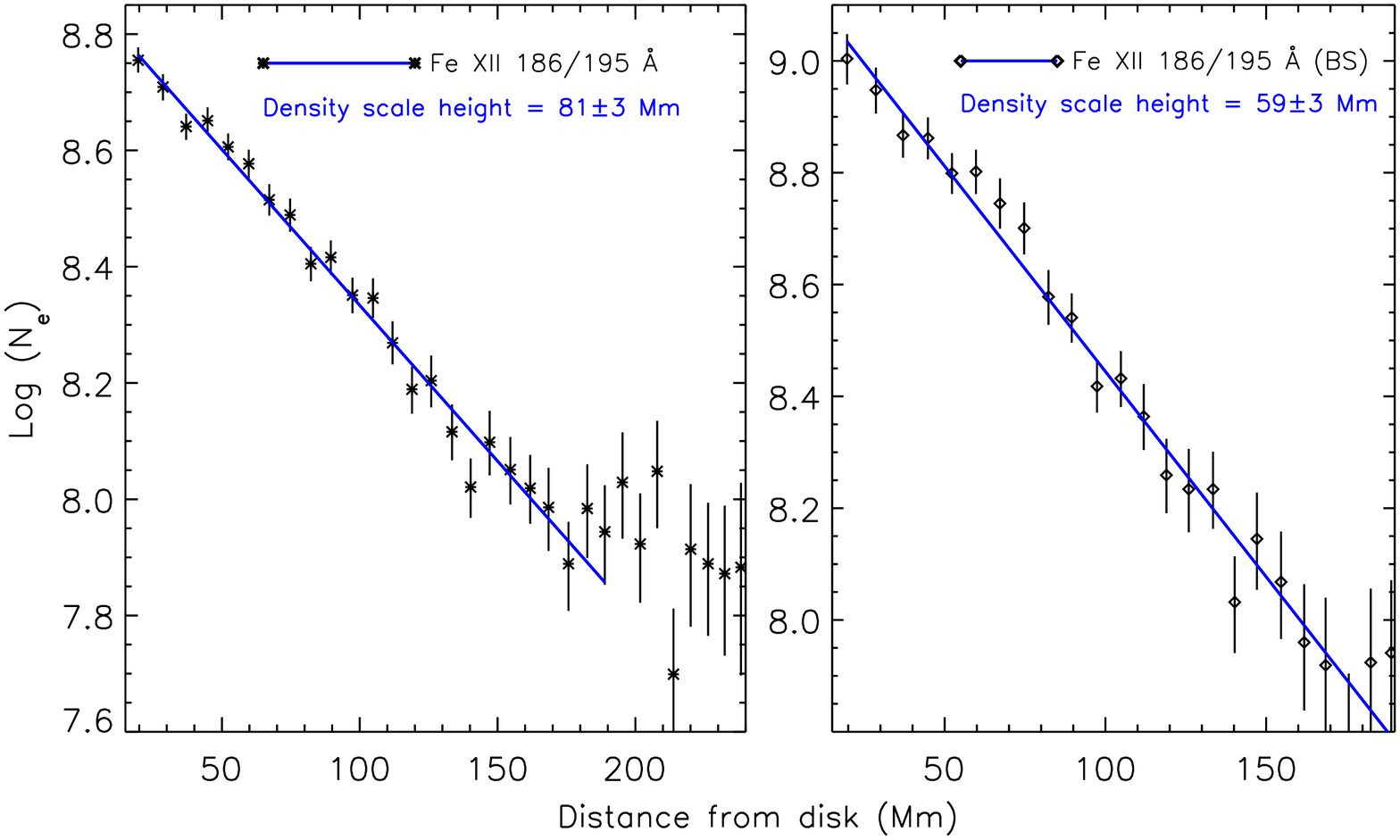}
\caption{Electron densities obtained from the line pair of Fe~{\sc xii} $\lambda 186.88/ \lambda 195.12$  with and without BS  plotted in right and left panels,  respectively. Densities are fitted with an exponential function
  to obtain the density scale height which are also printed on the figures.}
\label{fig:dens-hydro}
\end{figure}

We fitted the electron number density variation with height along the loop  with an exponential function $N_e=N_0 exp(-h/H_d)$ 
using MPFIT routines \citep{2009ASPC..411..251M}. The fit provides the density scale height $H_d$ along the coronal loop. 
For this purpose, we chose only the Fe~{\sc xii} line pair as density measurements from Si~{\sc x} line pair show relatively
larger error bars and are more noisy. We fitted both the density profiles obtained from  Fe~{\sc xii} line pair with and without
BS. Fitted profiles to density variations are plotted in Figure~\ref{fig:dens-hydro}. The fit resulted
in density scale heights of 59$\pm$3 and 81$\pm$3 Mm for density profiles with and without background subtracted, 
respectively. 

Recalling that the background subtracted densities are larger near the limb compared to without background subtracted densities, 
whereas both are almost the same in the far off-limb region. This indicates a rapid fall of density in the background subtracted
measurements, and thus gives the smaller density scale height as compare to without background subtracted measurements.
Density scale heights obtained from the fits are also labelled in the Figure~\ref{fig:dens-hydro}.

Electron number densities obtained along the loop can be compared with the hydrostatic equilibrium model
\citep[e.g.][]{1999ApJ...515..842A,2015ApJ...800..140G}. Electron number density profile in hydrostatic equilibrium is given by

\begin{equation}
 N_e(h)=N_e(0)exp \left(- \frac{h}{\lambda (T_e)} \right)
\label{eq:hydrostatic}
,\end{equation}

where $\lambda$ is the density scale height given by

\begin{equation}
 \lambda(T_e)= \frac{k_b T_e}{\mu m_H g} \approx 46 \left[ \frac{T_e}{1~MK} \right] [Mm]
\label{eq:scale}
,\end{equation}

where $k_b$ is the Boltzmann constant, $T_e$ is electron temperature, $\mu$ is  mean molecular weight
 ($\approx$ 0.7 for the solar corona), $m_H$ is mass of the hydrogen atom, and $g$ is acceleration due
 to gravity at the solar surface \citep[see e.g.][]{1999ApJ...515..842A}. 

The density scale height for hydrostatic equilibrium is calculated to be $\approx 63$ Mm for an electron temperature
of 1.37 MK as obtained from the EM-loci method. The observed density scale heights are 59$\pm$3 and 81$\pm$3 Mm for 
with and without background subtracted density profiles, respectively. These are reasonably close to the hydrostatic equilibrium height.

\begin{figure}[htbp]
\centering
\includegraphics[width=0.45\textwidth]{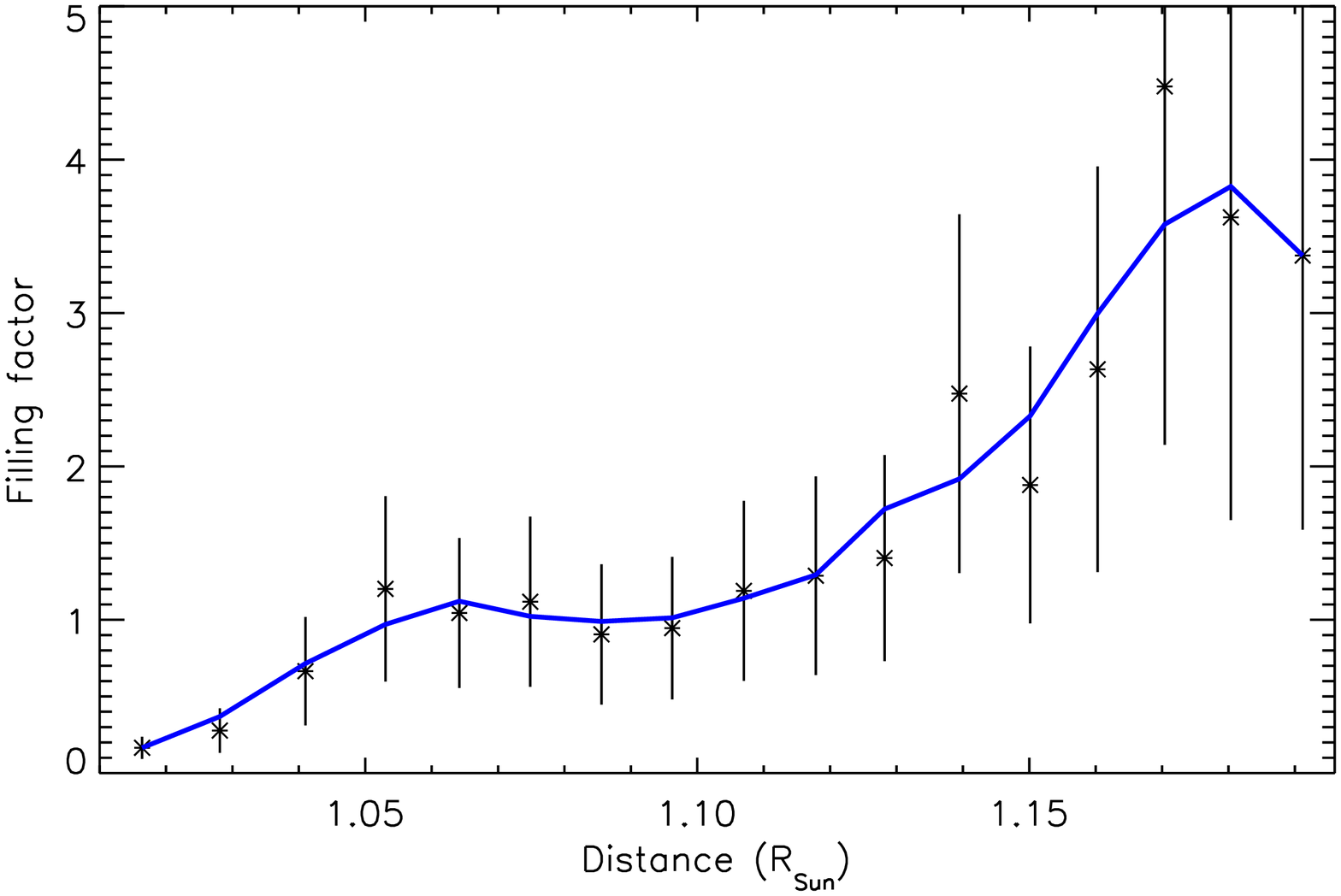}
\caption{Plasma filling factor obtained along the coronal loop.}
\label{fig:ff}
\end{figure}

 We also calculated coronal plasma filling factor, which is defined as the ratio of the volume emitting plasma 
 contributing to the EUV emission lines to the total volume observed.
 This parameter provides information on whether the observed coronal structures are resolved or unresolved with the 
 given instrument, using the formula  $ \phi=\frac{EM}{N_e^2 h}$,  where $EM$ is obtained in Section~\ref{sec:temp},
 N$_e$ is measured in Section~\ref{sec:dens}, and $h$ is column depth, which is assumed to be width of the loop
 measured in Section~\ref{sec:data}. The obtained plasma filling factors along the loop are plotted in Figure~\ref{fig:ff}.
 Error bars on filling factors were obtained following standard procedures, which include errors on N$_e$, EM,
 and loop width (1$\sigma$\ error on FWHM of geometric profile of loop). Near the limb, filling factors are less
 than unity and increase with height as 
  obtained in previous measurements \citep[e.g.][]{2009ApJ...694.1256T}. At some distance, the filling factor becomes
  more than unity, which is unrealistic. One interpretation is that we could be seeing an arcade of loops, where the 
  distance along the line of sight is greater than the measured geometric width.

\begin{figure*}[thbp]
\centering
\includegraphics[width=0.7\textwidth]{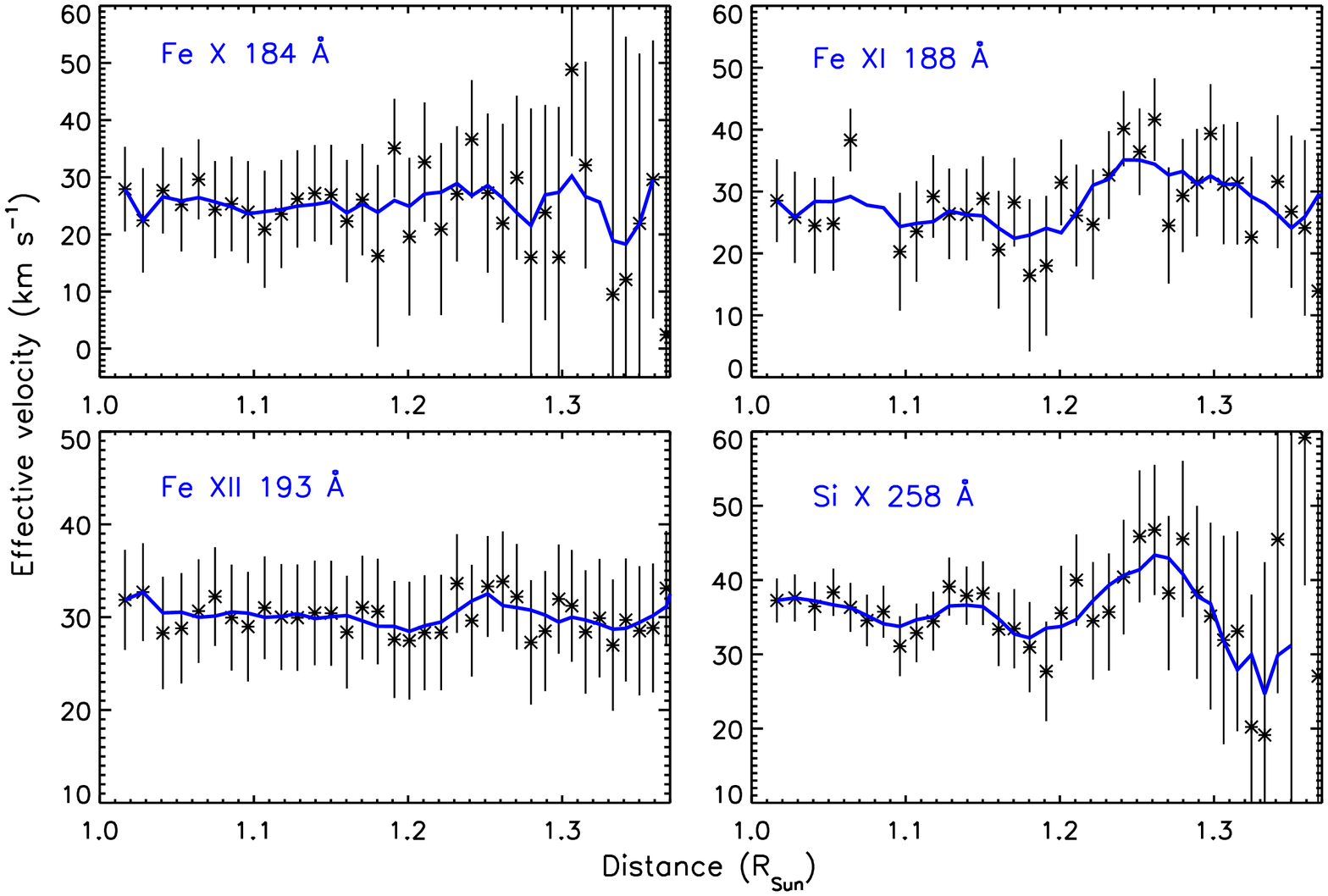}
\caption{Effective velocity (combined thermal and non-thermal velocity) obtained along the coronal loop from various 
spectral lines as labelled.}
\label{fig:evel}
\end{figure*}

\subsection{Effective and non-thermal velocity}
\label{sec:vel}

The observed FWHM of any coronal spectral line is given by

\begin{equation}
FWHM = \left[4 ln 2 \left(\frac{\lambda}{c}\right)^2 \left(\frac{2 k_b T_{i}}{M_{i}} +{\xi}^2\right) + W^{2}_{inst}\right]^{1/2}
\label{eq:fwhm}
,\end{equation}

where $T_{i}$ is ion temperature, $M_{i}$ is ion mass,  $\xi$ is non-thermal velocity, and $W_{inst}$ is the instrumental width.

After subtracting the instrumental width, the observed intrinsic line width of solar origin can then be expressed as
an effective width and velocity $v_{eff}$ \citep[e.g.][]{2011ApJ...741..107H},

\begin{equation}
v_{eff} =  \left(\frac{2 k_b T_{i}}{M_{i}} +{\xi}^2\right)^{1/2}
\label{eq:veff}
.\end{equation}

 Thus, $v_{eff}$ mainly depends upon $T_{i}$ and ${\xi}$. Upon assuming plasma to be in thermal equilibrium, $T_{i}$ 
 can be  approximated as $T_{e}$ (i.e. $T_{i} \approx T_{e}$), where $T_{e}$ is obtained in Section~\ref{sec:temp}. 
 $T_{i}=T_{e}$ is a common valid assumption which may not be true for coronal holes. We do not expect
 any large departures in this assumption for this relatively quiescent loop, at least in the main part of the loop.
 As an order of magnitude for a Fe~{\sc xii} ion, using the CHIANTI recombination rates and the measured 
electron densities, the recombination time near the limb turns out to be about 5 s, 
 while at the lower density of $10^8$ cm$^{-3}$  the value becomes almost 1 min,  which significantly
 validates our assumption. In Figure~\ref{fig:evel}, we plot the effective velocity obtained from Fe~{\sc x} 184.54,
Fe~{\sc xi} 188.23, Fe~{\sc xii} 193.51, and Si~{\sc x} 258.37 \AA\ spectral lines. 
Fe~{\sc xii} 193.51 \AA , which is the strongest line, shows an almost constant effective velocity with height.
 Other lines show a large scatter in data points in the far off-limb region but give a hint of slight increase in the velocity
 up to 1.3 R$_{\odot}$.

\begin{figure*}[htbp]
\centering
\includegraphics[width=0.7\textwidth]{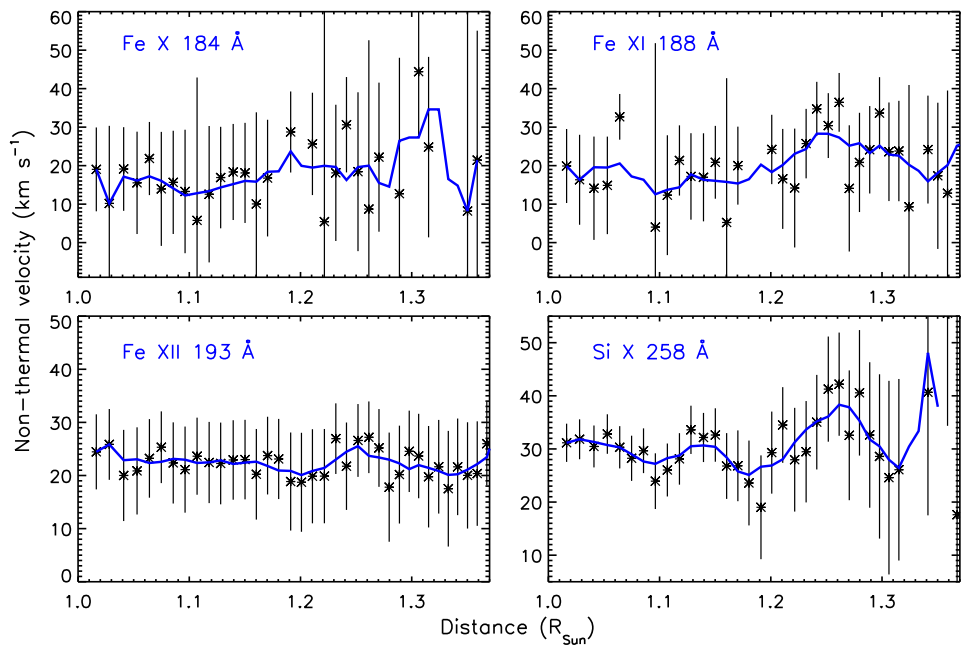}
\caption{Non-thermal velocity obtained along the coronal loop from various spectral lines as labelled.}
\label{fig:ntvel}
\end{figure*}

In Figure~\ref{fig:ntvel}, we plot the non-thermal velocities (${\xi}$) obtained from Fe~{\sc x} 184.54,
Fe~{\sc xi} 188.23, Fe~{\sc xii} 193.51, and Si~{\sc x} 258.37 \AA\ spectral lines. For this purpose,
we incorporated electron temperature $T_e$ obtained in Section~\ref{sec:temp} to calculate the thermal velocity.
Again Fe~{\sc xii} 193.51 \AA , the strongest line, shows almost constant non-thermal velocity with height.
Other lines show some increasing trend in the non-thermal velocity with height, however, the
data points are highly scattered. The average non-thermal velocity, as obtained from Fe~{\sc xii} 193.51 \AA\, is 
$\approx$23 km s$^{-1}$. Si~{\sc x} 258.37 \AA\ shows an average value of $\approx$30 km s$^{-1}$ with some increase
with height. However, non-thermal velocities from Fe~{\sc x} 184.54, and Fe~{\sc xi} 188.23 \AA\ show relatively
lower values. Interestingly, \citet{Delzanna2019} has noted that the instrumental width for Fe~{\sc xii} 193.5 \AA\ line
appears to be constantly larger than that of the other lines. They also noticed the overestimation in line widths
from Fe~{\sc xii} 193.51 \AA\ line, which is partly due to instrumental issues and partly due to opacity effects in
the stronger lines of Fe~{\sc xii} based on line ratio of Fe~{\sc xii} $\lambda193.51/\lambda195.12$. 
However, in our coronal loop, deviation of line ratio is within 10\% (see Figure~\ref{fig:int-rat}).
Thus, non-thermal velocity estimation from Fe~{\sc xii} 193.51 \AA\ is mildly affected. 
 

\subsection{Alfv\'en wave energy flux}
\label{sec:energy}

The Alfv\'en wave energy flux density is given by 

\begin{equation}
 E_D=2 \rho \xi^2 V_A=2 \sqrt{\frac{\rho}{4\pi}}~\xi^2~B
 \label{eq:energyd}
,\end{equation}

where $\rho$ is mass density ($\rho=m_p N_e$, $m_p$ is proton mass, and $N_e$ is electron number density), $\xi$ is Alfv\'en wave
velocity amplitude, 2 for two degrees of freedom, and $V_A$ is Alfv\'en wave propagation velocity given as $B/\sqrt{4\pi
  \rho}$. Therefore, the total Alfv\'en wave energy crossing a surface area A per unit time is given by

\begin{equation}
 E_A= \frac{2}{\sqrt{4\pi}}~\sqrt{m_p N_e}~\xi^2~B~A 
\label{eq:energyf}
.\end{equation}

\begin{figure}[htbp]
\centering
\includegraphics[width=0.45\textwidth]{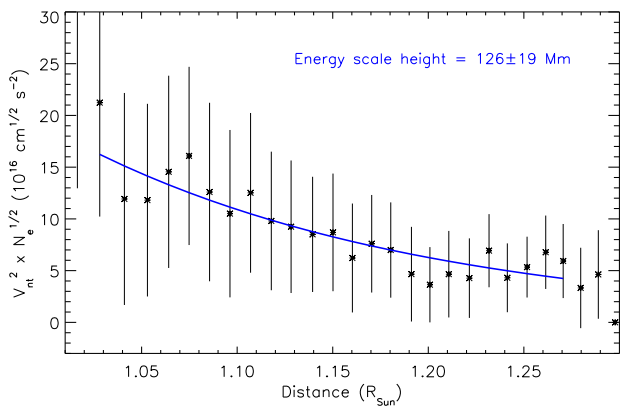}
\caption{Proportional Alfv\'en wave energy ($\alpha \sqrt{N_e}
  \xi^2$) obtained along the coronal loop from the Fe~{\sc xii} 193.51
  \AA\ spectral line. The wave energy is fitted with exponential
  function and the scale height obtained is also printed.}
\label{fig:alfven}
\end{figure}

Henceforth, the total Alfv\'en wave energy propagating along the loop depends on the electron number density, wave amplitude, 
magnetic field, and area of cross section.
However, the product of magnetic field $B$ and cross-section area $A$ is always a constant along the loop owing to
magnetic flux conservation. Therefore, the total Alfv\'en wave energy propagating along the loop is always proportional
to $\sqrt{N_e}~\xi^2 $.
The product of $\sqrt{N_e}~\xi^2 $ is expected to be constant with height if the total Alfv\'en wave energy is
conserved as wave propagates along the loop. Thus, we obtain the proportional Alfv\'en wave energy ($ \propto \sqrt{N_e}~\xi^2 $)
along the loop using the $N_e$ (see Section~\ref{sec:dens}) and $\xi$ (see Section~\ref{sec:vel}) obtained from Fe~{\sc xii} 
spectral lines. In Figure~\ref{fig:alfven}, we plot variations of $\sqrt{N_e}~\xi^2$ with height. The plot shows that
$\sqrt{N_e}~\xi^2$ decreases with height. Thus this provides evidence of damping of Alfv\'en wave energy with height
along the coronal loop, although associated error bars are very large on individual data points.

We further obtain the damping length using the equation \citep[e.g.][]{2017ApJ...836....4G}

\begin{equation}
F_{wt}\approx A~\sqrt{N_e}~\xi^2 e^{-h/D_l}
\label{eq:damping}
,\end{equation}

where $D_l$  is \textquoteleft damping length\textquoteright\ for the decay of total Alfv\'en wave energy with height,
and A is appropriate constant. Upon fitting with MPFIT routines \citep{2009ASPC..411..251M}, we found the damping length
to be 126$\pm$19 Mm. The damping length obtained by \citet{2017ApJ...836....4G} using Fe~{\sc xii} 192.39 \AA\ was also of
the same order in the active region. Thus, both the results are in good agreement.

\section{Discussion and summary}
\label{sec:discuss}

In this work, we presented the estimation of various plasma parameters along the long coronal loop using the EIS spectroscopic 
data. Spectroscopic diagnostics were used to measure electron density, temperature, non-thermal velocity, and thus Alfv\'en wave
energy flux along the loop length. The loop was clearly visible in the EIS Fe~{\sc x} 184.54, Fe~{\sc xi} 188.23, Fe~{\sc xii} 193.51,
and Si~{\sc x} 258.37 \AA\ spectral lines covering the peak formation temperature of 1.1-1.6 MK. The cross section (geometric 
width) of the loop increases with height. The observed width is about 20 Mm near the limb which increases up to 80 Mm at the distance
of 1.37 R$_\odot$. Increase in the loop cross section with height is also seen in the \citet{2015ApJ...800..140G}.
However, most of the coronal loops have been found to show roughly constant cross section or slight increase in cross section with
height \citep[e.g.][]{2000SoPh..193...77W}. It is interesting to note that no systematic study exists on the cross-section
variation of fan loops with height.

We obtained the electron density along the loop reaching far off-limb distance. Densities were obtained using the density sensitive 
EIS spectral line pairs Fe~{\sc xii} $\lambda 186.88/ \lambda 195.12$ and Si~{\sc x} $\lambda 258.37/ \lambda 261.04$.
Obtained number densities are around $10^9$ cm$^{-3}$ near the limb, which falls to $\approx 10^{7.9}$ cm$^{-3}$ at 
far off-limb distances. \citet{2003A&A...406.1089D} also obtained the electron densities along the off-limb coronal loops
and found a high density $\approx 3\times10^9$ cm$^{-3}$ near the base and rapid fall with height. 
We also compared the fall in density with the hydrostatic equilibrium model, which indicates that the loop is almost in hydrostatic equilibrium.
Obtained with and without BS, density scale heights are 59$\pm$3 and 81$\pm$3 Mm, respectively, which are
close to the hydrostatic scale height of 63 Mm. \citet{2014ApJ...780..177L} also studied the cool loops in off-limb region and found
density scale heights in the range 60-70 Mm from the Fe~{\sc xii} $\lambda 186.88/ \lambda 195.12$ line pair. 
Interestingly, \citet{2003ApJ...587..439W} found long, cool loops
to be overdense whereas short hot loops to be underdense on comparison with
the static solutions. However, the existence of such loops
may be explained by impulsive heating models \citep[e.g.][]{2008ApJ...682.1351K}.

We used the EM-loci technique to estimate
electron temperature along the coronal loop. Results obtained clearly
show that plasma along the line of sight in the loop is isothermal as all the EM curves cross at a single point.
The result also indicates that the temperature along the loop is also constant (within the obtained error bars). Thus, the observed loop
is isothermal across and along the loop with a temperature around $\approx 1.37$ MK. \citet{2009ApJ...694.1256T} also studied
the long coronal loop, however on-disc, and found the loop to be nearly isothermal along the line of sight. However, these authors noted that
temperature along the loop increases from 0.8 MK near the base to 1.5 MK at the height of 75 Mm.

We estimated the effective and non-thermal velocities along the coronal loop length. 
The strongest line Fe~{\sc xii} 193.51 \AA\ shows an almost constant effective and non-thermal velocity with height.
Other lines show a somewhat increasing trend with height, however, there is  large scatter present in the data points. 
The non-thermal velocity obtained from Fe~{\sc xii} 193.51 \AA\ line is around 23 km~s$^{-1}$. 
\citet{2017ApJ...836....4G} found an increase in non-thermal velocity from $\approx$ 24 km~s$^{-1}$ near the limb to 
$\approx$ 33 km~s$^{-1}$ around the height of 80 Mm along the coronal loop from Fe~{\sc xii} 192.39 \AA\ line. 
However, \citet{2014ApJ...780..177L} found decrease in non-thermal velocities along the cool
loop and dark lane with height in the off-limb active region using the Fe~{\sc xii} 195.12 \AA\ line.
These studies have associated the decrease in non-thermal velocity with height to damping of Alfv\'en waves. 
Various reports on the change in non-thermal velocity with height in the polar and equatorial region also exist,
which were associated with the propagation of Alfv\'en waves in the corona 
\citep[e.g.][]{1990ApJ...348L..77H,1998SoPh..181...91D,1998A&A...339..208B,2009A&A...501L..15B,2012ApJ...751..110B}.
In this study, however, it is surprising to see almost constant non-thermal velocity up to such a large distance.
One of the possibilities could be that as we are unable to see the foot-point of the loop, the initial increase of non-thermal
velocity with height observed in the previous studies \citep[e.g.][]{2017ApJ...836....4G} is hidden behind the limb. 

Recently, \citet{2017ApJ...849...46V} developed an Alfv\'en wave turbulence model for the heating of coronal loops.
Their model was able to reproduce a coronal loop with a temperature about 2.5 MK and predicted non-thermal velocity to be about
27 km~s$^{-1}$. However, in our observations of a coronal loop, the temperature is about 1.37 MK and non-thermal velocity is about 
$\approx$23 km~s$^{-1}$ as measured from  Fe~{\sc xii} 193.51 \AA\ spectral line. It is more likely that such coronal loop observations 
can easily be reproduced from Alfv\'en wave turbulence models \citep[e.g.][]{2014ApJ...786...28A}.
The recent Alfv\'en wave solar model of \citet{2017ApJ...845...98O} also provides good agreement between the predicted and observed
non-thermal broadening of spectral lines observed by SUMER/SOHO. These authors also modelled the electron temperature and density in the
pseudo-streamer and found these values to be consistent with the observations.

Off-limb studies of the quiet Sun have also provided little variation in the non-thermal broadening of line profiles up to about
1.3 R$_\odot$
\citep[e.g.][]{1998SoPh..181...91D,2005A&A...435..733W}. A recent study of \citet{Delzanna2019} of quiet-Sun region 
up to 1.5 R$_\odot$ had also shown almost no significant variation in the non-thermal velocty of Fe~{\sc xiii} 202 \AA\ spectral line
(around 15-20 km~s$^{-1}$).  However, they also noticed an overestimation in the line widths of the
Fe~{\sc xii} 193.51 \AA\ line. This is partly due to
instrumental issues and partly due to opacity effects in the stronger lines of Fe~{\sc xii} based on
line ratio of Fe~{\sc xii} $\lambda193.51/\lambda195.12$. In this
study, we also found that the Fe~{\sc xii} line ratio is not
constant at 0.67 as expected, however, deviation is within 10\%. Henceforth, line-width measurements would not have been greatly
affected in this case.

There are number of theoretical and observational studies which suggest that coronal loops are composed of thin unresolved
multi-strands which are heated impulsively \citep[see review by][]{2009ASPC..415..221K}. These impulsively heated loops lead
to field-aligned high-speed evaporative flows, which may lead to non-thermal broadening of spectral line profiles
\citep[e.g.][]{2006ApJ...647.1452P}. However as our study is pointed towards an off-limb region, the magnetic field lines
are generally directed radially outward, and thus would be orientated nearly perpendicular to the our line of sight.
Therefore, the contribution of Doppler velocities due to high-speed evaporative flows in the measurement of the non-thermal
velocities would be minimal. Also, the observed loop is found to be nearly in hydrostatic equilibrium, and thus there would be
hardly any plasma flows present along the loop. Therefore, the observed non-thermal velocity is most likely due to
the transverse Alfv\'en wave propagation along the loop structure. However, a small contribution due to some small component
of high-speed evaporative flows cannot be ruled out. 

We estimated the total Alfv\'en wave energy flux along the coronal loop. Although we do not find any significant decreasing
trend in non-thermal velocity with height, we find the damping of Alfv\'en waves with height based on the estimates carried out with
electron density and non-thermal velocity. The damping of Alfv\'en waves 
has also been found in different solar regions from the calculation of Alfv\'en wave energy flux using both electron density
and non-thermal velocity \citep[e.g.][]{2013ApJ...776...78H,2014ApJ...795..111H,2017ApJ...836....4G}.
\citet{2017ApJ...836....4G} obtained the damping length of Alfv\'en wave energy propagation to be 74--78 Mm for a coronal loop
observed up to the distance of 140--150 Mm, whereas in this study we
found the damping length to be around 125 Mm for a loop
observed up to the distance of 270 Mm. These results suggest that damping length of Alfv\'en waves may also depend
on the coronal loop length. However, only a statistical study over large sample of different loop lengths can confirm this.

\begin{figure}[thbp]
\centering
\includegraphics[width=0.48\textwidth]{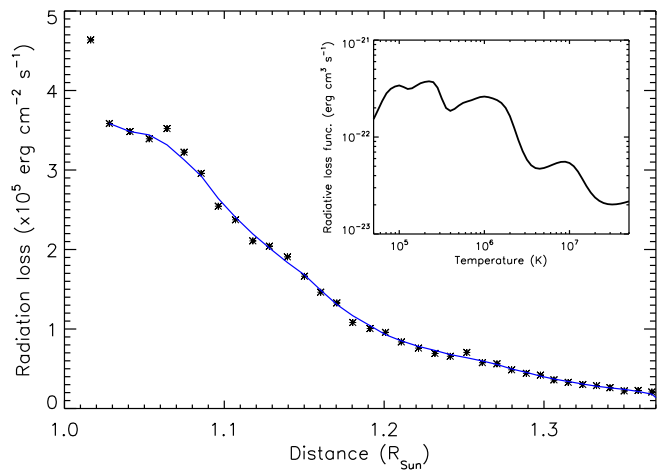}
\caption{Energy lost due to radiation with height along the loop length. Radiative loss function with temperature is also plotted
as obtained from CHIANTI v.8 database at electron number density of $10^{8.5}$ cm$^{-3}$.}
\label{fig:rloss}
\end{figure}

Upon assuming a magnetic field strength of the order of 10 G in an active region, \citep[e.g.][]{2000ApJ...541L..83L}, 
Alfv\'en wave energy fluxes are found to be decreasing from $\approx 1.5 \times 10^6$ erg cm$^{-2}$ s$^{-1}$ near the limb
to  $\approx 0.3 \times 10^6$ erg cm$^{-2}$ s$^{-1}$  at  around height of 1.2 R$_{\odot}$.
The estimated energy flux is less than the energy flux of the order of $10^7$  erg cm$^{-2}$ s$^{-1}$ required to
maintain the active region corona \citep{1977ARA&A..15..363W}. 
However, this energy requirement is obtained near the base of the corona. As we are higher up in the corona (loop foot-point being behind
the limb), the energy required to maintain the coronal losses would be smaller as density decreases with height. Thus, we
calculate the expected energy loss due to radiation from the loop using the radiative loss function and EM
obtained along the loop.
The associated radiative loss function is calculated using the CHIANTI v.8 database \citep{2015A&A...582A..56D} for the average electron density of 
$10^{8.5}$ cm$^{-3}$. For the purpose, we utilise solar coronal abundances provided by \citet{2012ApJ...755...33S}.
However, there may be an uncertainty of a factor of two in loss function just because of uncertainties in the abundance measurements.
We use the radiative loss function at an average temperature of 1.37 MK and EM obtained along the loop to
calculate the radiation loss along the loop length. The obtained energy lost from radiation and respective radiative loss function are plotted in 
Figure~\ref{fig:rloss}. The plot shows that energy lost from radiation is of the order of $10^5$  erg cm$^{-2}$ s$^{-1}$, which indicates that
Alfv\'en wave carries sufficient energy to maintain such coronal losses. However, these calculations provide just order of magnitude estimates
and should be taken with care.

In summary, we have found that the long coronal loop which we have studied is almost in hydrostatic equilibrium and is isothermal across
and along the loop. We have found evidence of damping of Alfv\'en waves which 
carries sufficient energy to support the energy lost by the loop due to radiation.

\begin{acknowledgements}
GRG acknowledges support from the UK Commonwealth Scholarship Commission via a Rutherford Fellowship during his stay in University of 
Cambridge, UK.
GDZ, and HEM acknowledge support by STFC (UK) via a consolidated grant to the solar/atomic physics group at DAMTP, University of
Cambridge. 

Hinode is a Japanese mission developed and launched by ISAS/JAXA, with
NAOJ as domestic partner and NASA and STFC (UK) as international partners.
It is operated by these agencies in co-operation with ESA and NSC (Norway).
CHIANTI is a collaborative project involving the University of Cambridge (UK),
George Mason University, and the University of Michigan (USA).

\end{acknowledgements}


\end{document}